\documentclass[prb,superscriptaddress, twocolumn]{revtex4-1}% APS journal style
\usepackage{mathtools,multirow,amssymb}
\usepackage[T1]{fontenc}
\usepackage{gensymb}
\usepackage{graphicx}

\usepackage{graphicx,booktabs}
\usepackage[verbose]{placeins}
\usepackage{blindtext}

\usepackage[colorlinks=true, urlcolor=black]{hyperref}

\usepackage{xcolor}

\begin{document}

\title{Point-defect engineering of MoN/TaN superlattice films: A first-principles and experimental study}

\author{Nikola Koutn\'a}
\affiliation{Institute of Materials Science and Technology, TU Wien, Getreidemarkt 9, A-1060 Vienna, Austria}
\affiliation{Department of Condensed Matter Physics, Faculty of Science, Masaryk University, Kotl{\' a}{\v r}sk{\' a} 2, CZ-611 37 Brno, Czech Republic}

\author{Rainer Hahn}
\affiliation{Institute of Materials Science and Technology, TU Wien, Getreidemarkt 9, A-1060 Vienna, Austria}

\author{Jakub Z\'{a}le\v{s}\'{a}k}
\affiliation{Erich Schmid Institute of Materials Science, Austrian Academy of Sciences, Jahnstrasse 12, A-8700 Leoben, Austria}

\author{Martin Fri\'{a}k}
\affiliation{Department of Condensed Matter Physics, Faculty of Science, Masaryk University, Kotl{\' a}{\v r}sk{\' a} 2, CZ-611 37 Brno, Czech Republic}
\affiliation{Institute of Physics of Materials, Academy of Sciences of the Czech Republic, \v{Z}i\v{z}kova 22, CZ-616 62 Brno, Czech Republic}
\affiliation{Central European Institute of Technology, CEITEC MU, Masaryk University, Kamenice 5, CZ-625 00 Brno, Czech Republic}

\author{Matthias Bartosik}
\affiliation{Institute of Materials Science and Technology, TU Wien, Getreidemarkt 9, A-1060 Vienna, Austria}

\author{Jozef Keckes}
\affiliation{Erich Schmid Institute of Materials Science, Austrian Academy of Sciences, Jahnstrasse 12, A-8700 Leoben, Austria}

\author{Mojm\'ir \v{S}ob}
\affiliation{Institute of Physics of Materials, Academy of Sciences of the Czech Republic, \v{Z}i\v{z}kova 22, CZ-616 62 Brno, Czech Republic}
\affiliation{Central European Institute of Technology, CEITEC MU, Masaryk University, Kamenice 5, CZ-625 00 Brno, Czech Republic}
\affiliation{Department of Chemistry, Faculty of Science, Masaryk University, Kotl{\' a}{\v r}sk{\' a} 2, CZ-611 37 Brno, Czech Republic}

\author{Paul H. Mayrhofer}
\affiliation{Institute of Materials Science and Technology, TU Wien, Getreidemarkt 9, A-1060 Vienna, Austria}

\author{David Holec}
\affiliation{Department of Materials Science, Montanuniversit\"{a}t Leoben, Franz-Josef-Strasse 18, Leoben A-8700, Austria}

\begin{abstract}
Superlattice architecture represents an effective strategy to improve performance of hard protective coatings. 
Our model system, MoN/TaN, combines materials well-known for their high ductility as well as a strong driving force for vacancies. 
In this work, we reveal and interpret peculiar structure-stability-elasticity relations for MoN/TaN combining modelling and experimental approaches.
Chemistry of the most stable structural variants depending on various deposition conditions is predicted by Density Functional Theory calculations using the concept of chemical potential.
Importantly, no stability region exists for the defect-free superlattice.
The X-ray Diffraction and Energy-dispersive $\text{X-ray}$ Spectroscopy experiments show that MoN/TaN superlattices consist of distorted fcc building blocks and contain non-metallic vacancies in MoN layers, which perfectly agrees with our theoretical model for these particular deposition conditions.
The vibrational spectra analysis together with the close overlap between the experimental indentation modulus and the calculated Young's modulus points towards MoN$_{0.5}$/TaN as the most likely chemistry of our coatings.
\end{abstract}

\maketitle

\section{Introduction}
Excellent mechanical properties of nitride-based protective coatings closely relate to their microstructure\cite{mayrhofer2006microstructural}.
To meet the demanding industrial requirements, coatings are nano-engineered as single or multilayer multicomponent systems. 
Especially superlattices (SLs), i.e, coherently grown nano-layers of two or more materials, represent a powerful concept to tune optical, magnetic, electronic, mechanical or tribologic properties \cite{stueber2009concepts,inspektor2014architecture,saha2016cross,wang2017systematic}.
Furthermore, SL architecture can enable formation of metastable phases, otherwise rather uneasy to synthesise experimentally.
The application higly relevant cubic AlN, for instance, was shown to be epitaxially stabilised in AlN/CrN\cite{Zhang2017-va,schlogl2013influence} or AlN/TiN\cite{koutna2019correlating,mei2004coherent,chawla2014effect} SLs.

In terms of mechanical properties, the great potential of superlattice architecture was demonstrated by \citet{helmersson1987growth} and \citet{barnett1998superhard}.
When the bilayer period of TiN/VN and TiN/NbN nanolayered coatings was set to 5 and 9\;nm, respectively, the hardness exceeded that of the monolithic film-forming phases by about 100\%.
Interface-induced enhancement of mechanical and/or tribologic performance beyond the limits of its individual components was further achieved for TiN/MoN, TiN/NbN, TiN/TaN, and TiN/CrN SL coatings \cite{nordin1998mechanical,zhang2015microstructure}.
\citet{hahn2016superlattice} further showed that fracture toughness and hardness of TiN/CrN exhibit almost the same dependence on the bilayer period, $\Lambda$, with the fracture toughness peak at $\Lambda\approx 6\;\text{nm}$ coinciding with the hardness peak. 

Additionally, an enormous versatility in structural and mechanical properties can be accomplished by intentionally using the typically unwanted products of Physical Vapour Deposition (PVD) processes: vacancies and point defects in general \cite{lao2017asymmetric,riedl2018influence,pacher2017vacancy,balasubramanian2016vacancy,glechner2018tuning,zhu2017effects}.
Recent work by \citet{buchinger2019toughness} particularly underlined the important role of theory-guided defect design, showing an impressive fracture toughness enhancement in TiN/WN$_x$ superlattices, 4.6\;MPa$\sqrt{\text{m}}$ for $\Lambda=10\;\text{nm}$, which presents one of the highest records among transition metal nitrides.
The authors proposed that the indentation modulus and fracture toughness dependence on $\Lambda$ (in particular, the emergence of the peak) relates to the changing vacancy content within WN layers. 
 
In this work, we chose molybdenum and tantalum nitride as the superlattice building materials.  
Not only do they exhibit many superior properties, such as excellent hardness and elastic moduli, high melting point, good thermal and electrical conductivity \cite{jauberteau2015molybdenum,bernoulli2013magnetron,liu2014fabrication,zhu2013phase}, but they also exist in many crystallographic and compositional variants \cite{klimashin2016impact,balasubramanian2017phase,li2014mechanical,koller2018structure}.
The usually experimentally desired phase is the cubic rocksalt (Fm$\overline{3}$m, $\#225$, rs). 
Importantly, the stability of this phase is conditioned by the presence of vacancies on either the metallic (MoN and TaN), or non-metallic (MoN) sublattice \cite{koutna2016point,ozsdolay2017cation,balasubramanian2017phase,stampfl2003metallic}.
Our {\it{ab initio}} pre-study \cite{koutna2018stability} indicated a strong potential for cubic-based MoN/TaN SLs, especially in terms of ductility. 
Yet the SL computational model was fully stoichiometric, i.e., disregarded the strong driving force of MoN and TaN for vacancies.
Such model somewhat surprisingly resulted in a phase transformation of the originally cubic SL building blocks towards tetragonally distorted $\zeta\text{-}$phases\cite{hu2017stabilization} (P4/nmm, $\# 129$).
Our reasoning was based on the enormous instability of the defect-free rocksalt structured MoN and TaN, which eliminate some of their soft phonon modes by lowering the symmetry.
Clearly, the presence of interfaces (and the related bi-axial coherency stresses) is essential for this phase transformation.

A highly intriguing question for the present study therefore is, whether the tetragonally distorted $\zeta\text{-}$phases can still compete with the probably more realistic model of the SL featuring cubic building blocks with vacancies.  
Furthermore, MoN/TaN SLs may generally prefer different vacancy contents and distribution than these known to work for the monolithic films. 
To provide a trustable basis for our modelling analyses, we employ series of experimental techniques. 
The main objective of our investigations is to propose a complete picture of atomic-scale architecture, thermodynamic stability, electronic and elastic properties of MoN/TaN SLs, which can further serve to formulate design guidelines for outstanding SL coatings.

\section{Methodology}

\subsection{Computational details}
The simulations were carried out within the framework of the Density Functional Theory (DFT) as implemented in the Vienna Ab-initio Simulation Package (VASP) \cite{Kresse1996Efficient, Kresse1999From} together with plane-wave projector augmented wave (PAW) pseudopotentials \cite{Kohn1965Self}.
In order to treat the exchange and correlation effects, we applied the Perdew-Burke-Ernzerhof generalized gradient approximation \cite{PhysRevLett.77.3865}.
The plane-wave cut\-off energy was set to 700\,eV, while the reciprocal space was sampled with $\Gamma$-centered Monkhorst-Pack meshes\cite{monkhorst1976special} equivalent to the product of the number of k-points and the number of atoms equal at least 25 000.

Our model of the defect-free MoN/TaN superlattice with the MoN-to-TaN molar ratio 1:1 was based on the conventional 8-atom cubic fcc cell. 
Applying the periodic boundary conditions, the 1$\times$1$\times (2n)$, $n=1,2,\ldots$, geometry produced SLs with $(001)$ interfaces and bilayer periods $\Lambda\approx 9n\;\text{\AA}$.
As shown in our pre-study\cite{koutna2018stability}, a full relaxation of such SLs breaks the cubic symmetry of the fcc building blocks and induces a structural transformation towards the tetragonal $\zeta\text{-phase}$ (P4/nmm, $\# 129$).
While the $\zeta\text{-TaN}$ is dynamically stable (unlike rs-TaN), the $\zeta\text{-MoN}$ still yields soft phonon modes.
Following these finally leads to the dynamically stable $\omega\text{-MoN}$ (P2$_1$/m, $\#11$).
The fcc, $\zeta$, and the $\omega$ structures of MoN and TaN are compared in Tab.~\ref{Tab: zeta vs rs}.

\begin{table*}[h!t!]
\caption{Lattice parameters, $a$ and $c$ (in \AA), formation energies, $E_f$ (in eV/at.), and stability of the fcc (Fm$\overline{3}$m), $\zeta$ (P4/nmm), and the $\omega$ (P2$_1$/m) structural variants.}
\centering
\begin{tabular}{cccccc|cccccc}
\hline
\hline
\multicolumn{6}{c}{MoN} &
\multicolumn{6}{c}{TaN} \tabularnewline
Phase & $a$ & $c$ & $E_f$ & Mech. stability & Dyn. stability & Phase & $a$ & $c$ & $E_f$ & Mech. stability & Dyn. stability \\
\hline
fcc-MoN & 4.364 & & $-0.008$ & NO & NO & fcc-TaN & 4.427 & & $-0.887$ & YES & NO \\
$\zeta$-MoN  & 4.248 & 4.544 & $-0.178$ & NO & NO & $\zeta$-TaN  & 4.202 & 5.119 & $-0.982$ & YES & YES \\
$\omega$-MoN  & 4.424 & 4.298 & $-0.245$ & YES & NO \\
\hline
\hline
\end{tabular}
\label{Tab: zeta vs rs}
\end{table*}

In this work, we studied the effect of vacancies in the already pre-relaxed SLs, i.e., composed of the $\zeta\text{-phases}$.
To keep the simulations computationally affordable, the supercell size was set to $2\times2\times4$ (128 atoms), leading to the bilayer period $\Lambda\approx1.91\;\text{nm}$ (after relaxation).
As we wanted to analyse the effect of interfaces themselves, a desired number of Mo, Ta, or N vacancies was randomly distributed at each (001) plane.
In particular, different vacancy configurations in terms of vacancy distribution within the layers ware produced even for the same vacancy content and the same type of missing atom.
Vacancies in the monolithic $\zeta\text{-MoN}$ and $\zeta\text{-TaN}$ were distributed employing the Special Quasi-random Structure (SQS) method \cite{Wei1990Electronic}. 

Despite both $\zeta$-phases and SLs had the overall tetragonal symmetry, the presence of vacancies generally led to 3 distinct lattices parameters, $a_1$, $a_2$, and $c$, after a full relaxation. 
The tetragonal lattice parameter $a$ of $\zeta\text{-Me}_{x}\text{N}_y$, Me$=$Mo, Ta, was calculated by averaging $a_1$ and $a_2$. 
In the case of SLs, such average represented an effective in-plane lattice constant (in the interface), while $c$ (in the direction perpendicular to the interface) represented the bilayer period, $\Lambda$. 
To quantify the effect of vacancies on each of the SL-building materials, we estimated an effective lattice parameter $c_{\text{eff}}^{\text{MoN}}$, $c_{\text{eff}}^{\text{TaN}}$, in MoN and TaN layers separately.
This was done by measuring interplanar distances between all pairs of neighbouring (001) planes (the $z$-coordinates of these planes were averaged fully relaxed $z$-coordinates of all ions occupying the same plane, i.e. with the same $z$-coordinate before the relaxation). 
Furthermore, to reflect that SLs with the same vacancy concentration but different vacancy distribution may differ in energies, we employed a weighted average
\begin{equation}
\overline{x}=\frac{\sum_{i=1}^nw_ix_i}{\sum_{i=1}^nw_i},
\end{equation}   
where $x_i$ is a structural parameter of interest (i.e., $\Lambda$, $a$ in the interface) and $w_i$ is the corresponding weight.
Index $n$ corresponds to the number of investigated distributions of the same vacancy type and overall concentration.
Denoting $E_{\text{tot}}^{\text{min}}$ the minimal formation energy of a SL with a specific vacancy type and concentration, the weight of a SL configuration with energy $E_{\text{tot},i}$ was calculated as
\begin{equation}
w_i=E_{\text{tot},i}/E_{\text{tot}}^{\text{min}},
\end{equation}
where $E_{\text{tot},i}, E_{\text{tot}}^{\text{min}}<0$ and $E_{\text{tot}}^{\text{min}}\neq 0$ (the wights are non-negative).

Thermodynamic stability of SL as well as monolithic phases was quantified with the energy of formation, $E_f$, calculated according to
\begin{equation}
E_f=\frac{1}{\sum_s n_s}\bigg(E_{\text{tot}}-\sum_s n_s\mu_s\bigg)\ ,
\label{Eq: Ef}
\end{equation} 
where $E_{\mathrm{tot}}$ is the total energy of the supercell, $n_s$ and $\mu_s$  are the number of atoms and the chemical potential, respectively, of a species $s$. 
Depending on the specific experimental conditions, the chemical potentials $\mu_\text{N}$, $\mu_\text{Mo}$ and $\mu_\text{Ta}$ may in principle reach any value below the upper limit given by the total energy (per atom) of bcc-Mo, $\mu_{\text{Mo}}(\text{bcc-Mo})$, bcc-Ta, $\mu_{\text{Ta}}(\text{bcc-Ta})$, and a half of the N$_2$ molecule, $\mu_{\text{N}}(\text{N}_2)$, respectively\cite{koutna2016point}.
These upper limits are conventionally referred to as the N-rich, Mo-rich and Ta-rich conditions, respectively.

The stress-strain method \cite{le2002symmetry, le2001symmetry, Yu2010-vr} was applied to calculate the fourth-order elasticity tensors $\mathbb{C}$ of selected systems according to the Hooke's law, $\sigma=\mathbb{C}\varepsilon$, which establishes a linear relation between stress, $\sigma$, and strain, $\varepsilon$.
Using the Voigt's notation, we conventionally transform the calculated fourth-order tensors, $\mathbb{C}$, to a symmetric $6\times6$ elastic constants matrix $\{C_{ij}\}$.
A positive definiteness of the elastic matrix, or equivalently, a positivity of its minimal eigenvalue, $\lambda_{\text{min}}$, served as a necessary and sufficient criterion for a mechanical stability of the corresponding structure\cite{mouhat2014necessary} provided other instabilities (e.g., soft phonon modes or magnetic spin arrangement) do not come forth. 
In principle, analysis of the phonon spectrum of a crystal is required to ensure the vibrational stability of the investigated material.
Here, the phonon spectra were calculated using the Phonopy package \cite{togo2008first}. 

The polycrystalline bulk, $B$, and shear, $G$, moduli were represented with the Hill's average\cite{hill1952elastic} of the upper Reuss's limit\cite{reuss1929berechnung} and the lower Voigt's bound \cite{voigt1928lehrbuch}. 
The Young's modulus was evaluated as
\begin{equation}
E=\frac{9BG}{3B+G},
\end{equation}
using the Hill's averages of $B$ and $G$.
The relative tendency for brittleness/ductility was predicted based on the Pugh's ratio, $B/G$, and Poisson's ratio, 
\begin{equation}
\nu=\frac{3B-2G}{6B-2G}.
\end{equation}
Values $B/G>1.75$\cite{pugh1954xcii} and $\nu>0.33$\cite{frantsevich1983elastic}, respectively, were conventionally interpreted as a sign for ductile behaviour.

\subsection{Experimental setup}
MoN/TaN SL films were deposited using the AJA International Orion 5 lab scaled deposition plant, equipped with two 2" unbalanced magnetron cathodes.
The base pressure of the sputtering chamber prior to the deposition was lower than $10^{-4}$\;Pa. 
Prior to the depositions, Si(100), $20\times 7 \times 0.38\;\text{mm}^3$, and MgO(100), $10\times 10 \times 0.5\;\text{mm}^3$, substrates were ultrasonically cleaned in acetone and ethanol for 5\;min each.
Subsequently, they were thermally cleaned at 500\;$\degree$C for 20\;min inside the chamber followed by etching for 10\;min (by applying constant $-750\;\text{V}$ bias voltage at an Ar pressure of 6\;Pa).

The depositions were carried out at a substrate temperature of 500\;$\degree$C in an Ar/N$_2$ gas mixture (7 sccm Ar and 3 sccm N$_2$) at a total pressure of 0.4\;Pa.
The current densities of the dc powered Mo and Ta targets (both 99.6$\%$ purity, Plansee Composite Materials GmbH) were set to 19.7\;mA/cm$^2$. 
To achieve a dense coating morphology, we applied a negative bias voltage of $-40\;\text{V}$\cite{mattox1989particle}.
Nanolayered structures were realized using a computer-controlled shutter system. 
The bilayer periods were nominally set to $\Lambda=1.5$, 3, and 6~nm, while the total number of bilayers were chosen to produce a film thickness of $\approx2\;\mu\text{m}$.

The X-ray diffraction patterns in Bragg-Brentano (BB) configuration were collected using a PANalytical XPert Pro MPD ($\theta\text{-}\theta$ diffractometer) equipped with a Cu-K$_{\alpha}$ radiation source ($\lambda=0.15418\;\text{nm}$).
The gracing incidence (GI) patterns were recorded with an incidence angle of 2.0$\degree$ using an Empyrean PANalytical ($\theta\text{-}\theta$ diffractometer) with a Cu-K$_{\alpha}$ source.

A coating with $\Lambda$ varying from 1.5 to 15\;nm was used for Transmission Electron Microscopy (TEM) studies.
A cross-sectional electron transparent lamella was prepared using a FEI Helios NanoLab G3 UC dual beam focused ion beam (FIB) microscope. 
The FIB was operated at accelerating voltages from 30\;kV to 1\;kV and currents from 20\;nA to 40\;pA. 
Cross-sectional scanning transmission electron microscopy (STEM) and energy dispersive spectroscopy (EDX) was performed using FEI Titan Themis equipped with a SUPER-X EDX detector. 
The microscope was operated at accelerating voltage of 300\;kV and probe current of 0.7\;nA.
The collected EDX signal was treated using TEAM software applying built-in standards.

Indentation hardness of our coatings was measured with the Fischer Cripps Laboratories ultra-micro indentation system (UMIS) equipped with a Berkovich diamond tip. 
A series of indentations with different loads (3--45\;mN) was performed for each multilayer system and substrate. 
Indentations with a total indentation depth larger than 10\% of the film thickness were ruled out, the rest of the data was evaluated following Oliver and Pharr\cite{pharr1992measurement}.
The indentation modulus was calculated by extrapolating a fit to zero indentation depth in order to minimise substrate effects\cite{fischer2006critical}.
The residual stress values were evaluated using Stoney's equation\cite{stoney1909tension}, whereby the curvature was determined employing optical profilometry.

\section{Results and discussion}

\subsection{Energetics and structure}\label{sectionA}

\subsubsection{Formation energies and structural parameters from first-principles}\label{Sec1,ab initio}
Prior to experimental investigations, formation energies, $E_f$, and lattice constants of MoN/TaN SL are carefully analysed depending on the defect type, concentration, and distribution.
As already these three parameters may draw a rather complex picture, we limit ourselves to the 128-atom simulation cell with $\Lambda\approx1.91\;\text{nm}$, i.e., do not bring another variable via altering the bilayer period.
To identify the most stable structural variants depending on specific deposition conditions, $E_f$ is evaluated as a function of metal and nitrogen chemical potentials, quantifying the availability of the respective element in experiment.
For example, the nitrogen chemical potential can be easily interpreted as the applied $N_2$ partial pressure.

\begin{figure*}[t]
	\centering
    \includegraphics[width=13cm]{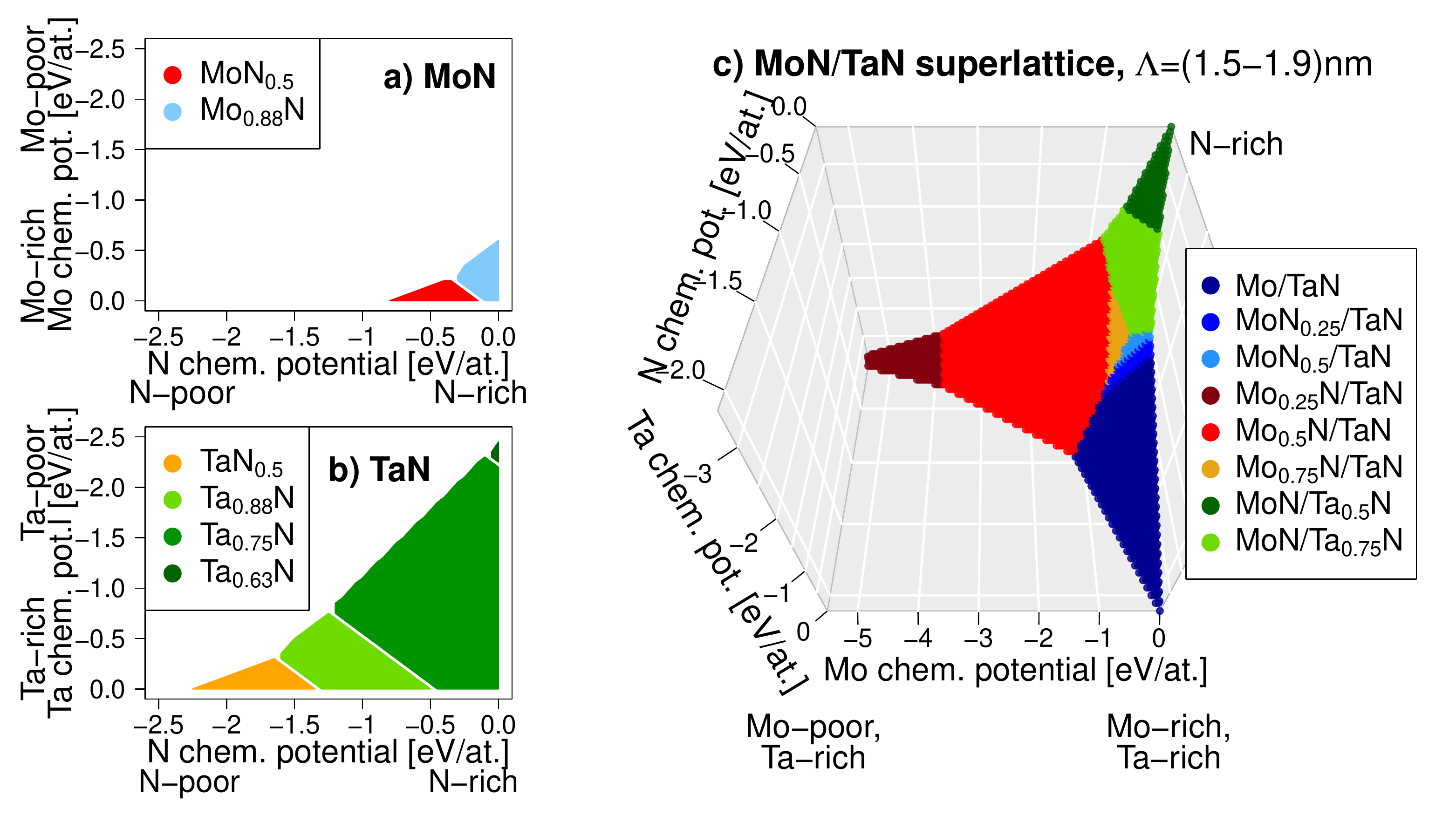}
	\caption{Phase diagram presenting the stable compositions of the $\zeta\text{-MeN}$, Me$=$Mo, Ta (a, b), and the MoN/TaN SL (c) as a function of the N, Ta and Mo chemical potentials. The uncoloured area corresponds to unstable compositions.}
\label{FIG: Ef}
\end{figure*}

To highlight the impact of interfaces, Fig.~\ref{FIG: Ef} contrasts the energetic trends for SLs with those obtained for their single-phase $\zeta\text{-Me}_{x}\text{N}_y$ components (Me$=$Mo, Ta).
A structure is deemed unstable for a combination of $\mu_\text{N}$, $\mu_\text{Mo}$, and $\mu_\text{Ta}$ yielding a positive value of $E_f$ (cf. Eq.\eqref{Eq: Ef}).
Nevertheless, a negative sign of $E_f$ does not yet imply mechanical and/or vibrational stability (investigated later in Sec.~\ref{B:Elasticity} and \ref{C:Phonons}).

Fig.~\ref{FIG: Ef} clearly demonstrates the important stabilisation role of vacancies, since neither the binary nor the SL systems show a stability region for a defect-free structure.
Regardless the values of chemical potentials, $E_f$ of the defect-free variant exceeds $E_f$ of the most favourable vacancy-featuring structure by about 0.1--0.2\;eV/at.

Starting with the monolithic binaries, the Mo--N system favours both vacancy types, metallic and nitrogen. 
The overall lowest $E_f$ is obtained for $\zeta\text{-MoN}_{0.5}$ and $\zeta\text{-Mo}_{0.88}\text{N}$.
Nevertheless, these polymorphs are stable in a rather narrow range of chemical potentials, suggesting that very specific experimental conditions are required for their synthesis.
The $\zeta\text{-TaN}$ with Ta or N substoichiometry shows a more extended stability region dominated by Ta$_{0.75}\text{N}$ and Ta$_{0.88}\text{N}$.
N vacancies in $\zeta\text{-TaN}$ are favoured when approaching the N-poor conditions, i.e., low N$_{2}$ partial pressures.
Such findings are similar to those predicted for rs-Mo$_x$N$_y$ and rs-Ta$_x$N$_y$\cite{koutna2016point,klimashin2016impact}. 

Despite clear similarities, the energetics of MoN/TaN SL is not a simple addition of the trends for the monolithic components.
Defected configurations with vacancies in either Mo or N sublattice of MoN, i.e., Mo$_{0.25\text{--}0.75}$N/TaN, MoN$_{0.25\text{--}0.5}$N/TaN, and Mo/TaN, dominate the stability diagram.
Consequently, coatings with highly defected MoN layers should be most likely produced experimentally.
The enormously high vacancy contents, i.e., above 50\%, may lead to significant structural changes.
Indeed, a close examination of MoN$_{0.25}$/TaN SL reveals that the MoN layers contain a mixture of two phases, $\zeta\text{-MoN}_y$ and bcc-Mo.
Going to the extreme case, the Mo/TaN SL is formed by a 45$\degree$-rotated bcc-Mo and $\zeta\text{-TaN}$ layers, i.e., [100]$_{\text{TaN}}\parallel [1\overline{1}0]_{\text{bcc-Mo}}$ (similarly to rs-CrN in Ref.\cite{zhang2013insights}). 
N-rich, Mo-rich and Ta-poor conditions enable stabilisation of Ta-deficient MoN/Ta$_{0.5\text{--}0.75}$N. 
In contrast to the monolithic TaN, N vacancies in TaN layers of the superlattice are highly unfavourable for all combinations of chemical potentials.

\begin{figure}[h]
	\centering
    \includegraphics[width=7.5cm]{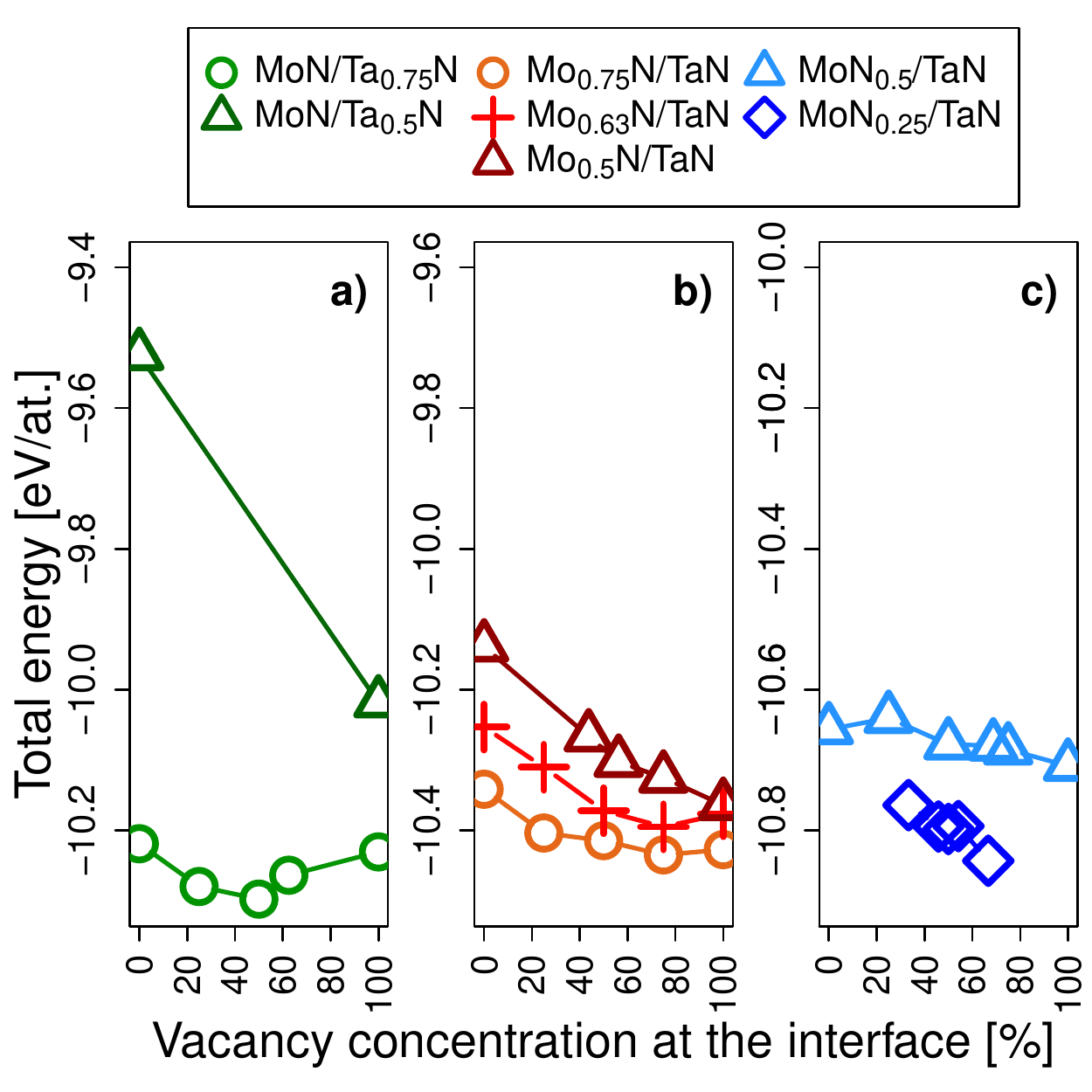}
	\caption{Total energy of the most stable SLs (cf. Fig.~\ref{FIG: Ef}) as a function of the vacancy concentration directly at interfaces. Panels (a, b, c) depict data points for Ta, Mo and N vacancies, respectively.}
\label{FIG: defect distribution}
\end{figure}

Not just the amount of vacancies, but also their spatial distribution, i.e., relative content at interfaces, might play an important role.
Fig.~\ref{FIG: defect distribution} provides an elementary insight into the relative segregation tendency of vacancies based on total energies, $E_{\text{tot}}$, which are independent of chemical potentials. 
The $E_{\text{tot}}$ minimum for MoN/Ta$_{0.75}$N SL is obtained when half of all vacancies accumulates at interfaces. 
Interfaces become even more energetically favourable when Ta vacancy concentration rises up to 50\;at.\%.
Mo vacancies in Mo$_{0.5\text{--}0.75}$N/TaN SL also tend to accumulate at interfaces. 
Specifically, the strongest preference of Mo vacancies for interfacial sites is predicted for Mo$_{0.5}$N/TaN: the $E_{\text{tot}}$ difference between the SL with all vacancies occupying (i) the interface and (ii) the ``bulk-like'' layers is $\approx 0.2\;\text{eV/at}$. 
Comparable trends are predicted for 50--75\;at.\% N vacancies in MoN layers.

\begin{figure*}[t!]
	\centering
    \includegraphics[width=14cm]{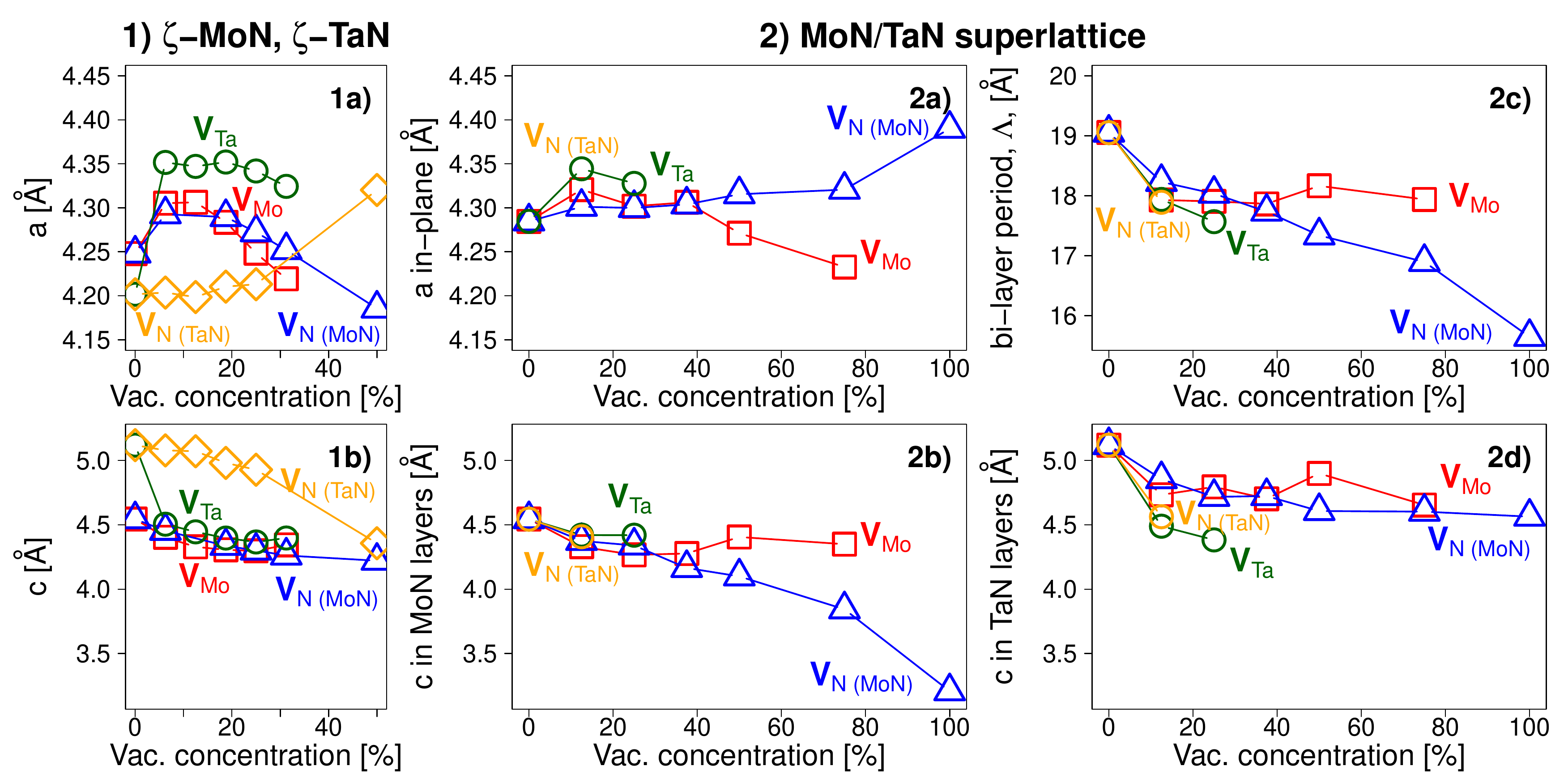}
	\caption{The impact of Mo, Ta, and N vacancies on the structural parameters: (1a--b) present lattice parameters for the monolithic $\zeta\text{-MeN}$, Me$=$Mo, Ta, while (2a--d) show data points for the MoN/TaN superlattices.}
\label{FIG: lattice parameters}
\end{figure*}

We will now proceed to a careful structural analysis, technical details of which are explained in the Methodology section.
Fig.~\ref{FIG: lattice parameters} suggests that the increase of vacancy content in $\zeta\text{-MoN}$ and $\zeta\text{-TaN}$ leads to a notable decrease of the tetragonal lattice parameter $c$.
This is compensated by the increase of $a$, with the only exception of N vacancies in $\zeta\text{-TaN}$.
For example, as a result of $6\%$-Ta-substoichiometry in $\zeta\text{-TaN}$, $c$ drops abruptly from 5.12 to 4.50\;\AA, accompanied by an enlargement of $a$ from 4.20 to 4.35\;\AA.
The simulation cells hence become more cubic as $a$ and $c$ get closer, which perfectly explains the comparable energetic trends for the defected cubic (cf.~Ref.~\cite{koutna2016point}) and the $\zeta\text{-phase}$ (Fig.~\ref{FIG: Ef}).   
Unlike that, the N vacancies in $\zeta\text{-TaN}$ do not break the tetragonal symmetry, as the $c/a$ ratio remains roughly constant ($\sim$1.2), independently of the defect content.
Analogically, this may explain different stability regions of $\zeta\text{-TaN}_{0.5}$ as compared to its cubic rocksalt counterpart.

Surprisingly, the in-plane lattice constant of the MoN/TaN SL ($\sim 4.28$\;\AA) exceeds $a$ of both the $\zeta\text{-TaN}$ ($\sim 4.20$\;\AA) and the $\zeta\text{-MoN}$ ($\sim 4.25$\;\AA).
We speculate that the unstable rs-MoN wants to eliminate its imaginary phonon frequencies by relaxation towards the monoclinic $\omega\text{-MoN}$ ($a\sim 4.42\;$\AA, cf.~Tab.~\ref{Tab: zeta vs rs}), i.e., not only to the tetragonal $\zeta\text{-MoN}$, which is still vibrationally unstable.
At the same time, the strong $\zeta\text{-TaN}$ dictates the SL tetragonal symmetry, allowing only for a partial cubic-to-$\zeta$ ($\omega$) phase transformation. 
Vacancies in MoN/TaN SLs decrease $\Lambda$ as compared with the reference value 19.1\;\AA\;for the defect-free SL.
This is in most cases accompanied by an expansion of the in-plane lattice constant. 
Fig.~\ref{FIG: lattice parameters} (2b, d) further suggests that vacancies in one of the materials decrease the effective lattice parameter $c$ in the corresponding (defected) sublattice, and---though not that significantly---also in the defect-free sublattice.
As this is accompanied by an increase of the in-plane lattice constant, we propose that vacancies stimulate a $\zeta$-to-fcc transition in defected layers.
Such effect might be partially transferred through the interface via the in-plane tensile stresses, explaining a decrease of the effective $c$ of the defect-free layers.

\subsubsection{Experimental determination of structure and defect content}
Our DFT calculations suggested that vacancies in MoN/TaN SL coatings are highly expectable, in particular in MoN layers. 
Rather specific deposition conditions may lead to Ta vacancies, while the defect-free system should be energetically very unfavourable.
Moreover, the highly defected layer material should be cubic, while the (nearly) perfect layer material should present tetragonal-like distortions.

\begin{figure}[h!]
	\centering
    \includegraphics[width=8.5cm]{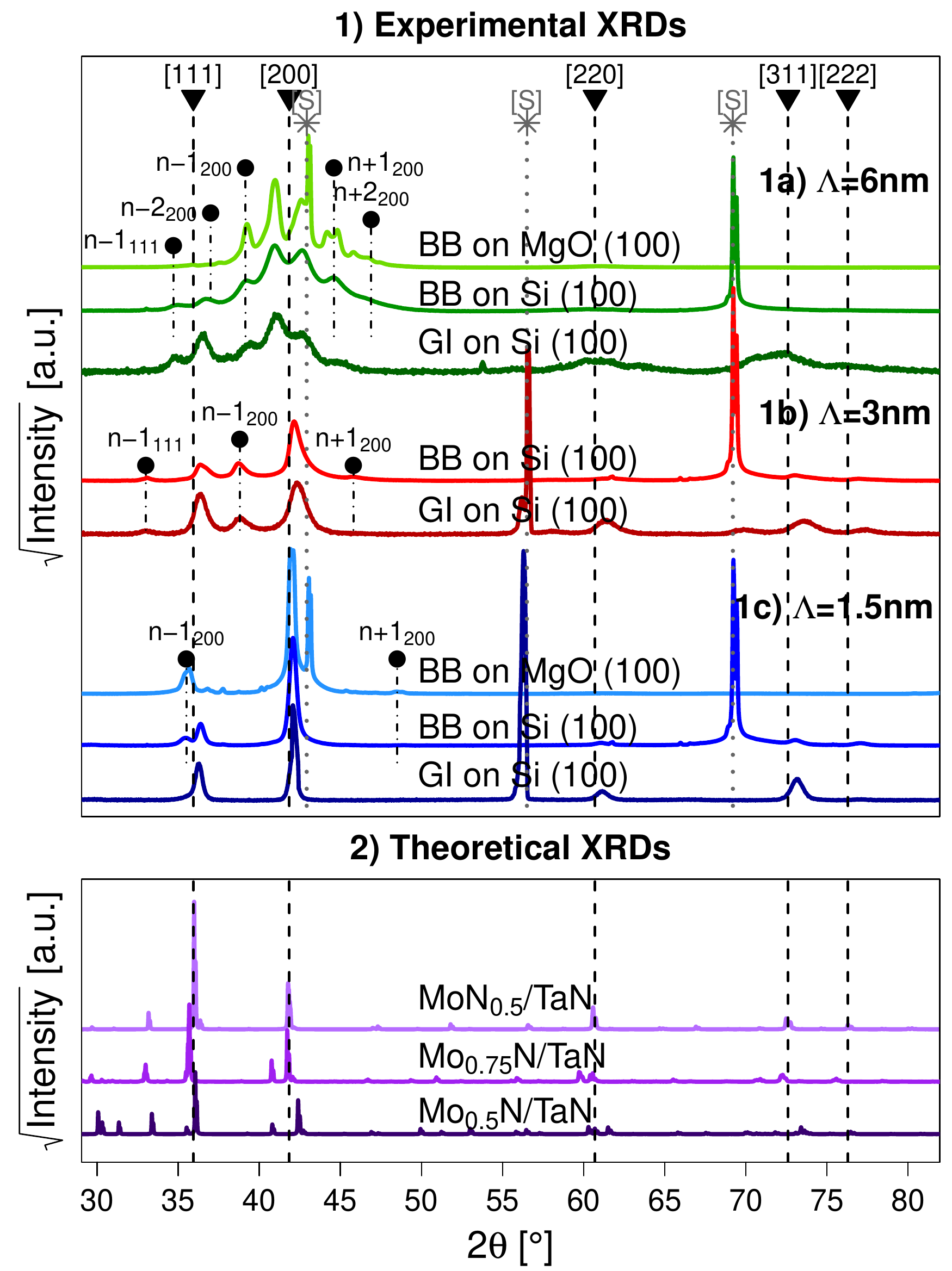}
    \caption{XRD patterns recorded in Bragg Brentano (BB) and Gracing Incidence (GI) configurations, of MoN/TaN thin films with $\Lambda=1.5$\;nm (1a), 3\;nm (1b), and 6\;nm (1c). Superlattice and substrate peaks are marked with triangles and stars, respectively. The double-peaks for $\Lambda=6\;\text{nm}$ correspond to the respective constituent ($2\theta_{\text{TaN}}<2\theta_{\text{MoN}}$). XRD patterns for the DFT-predicted MoN$_{0.5}$/TaN, Mo$_{0.75}$N/TaN, and Mo$_{0.5}$N/TaN superlattices (2).}
\label{FIG: XRD}
\end{figure}

In order to verify these predictions, we deposited MoN/TaN coatings with nominal bilayer periods, $\Lambda_{\text{nom}}$, of 1.5, 3 and 6\;nm.
According to the XRD patterns (Fig. \ref{FIG: XRD}), the coatings exhibit a distorted cubic SL structure with sharp interfaces and a strongly oriented (200) texture (cf.~BB measurements on coated Si substrates).
The measurements of coatings on MgO (100) show no signs of peaks other than (200) due to the strong template effect originating from the similar lattice parameters of MoN/TaN SLs and MgO ($a=4.21$\;\AA).
The MgO substrate peak (42.92$\degree$) in the BB configuration belongs to the (200) planes, while the Si substrate peak (69.13$\degree$) belongs to (400) planes. 
Based on the positions of the satellite peaks, $\theta_\pm$, and the position of the main peak, $\theta_B$, we directly calculate the bilayer period, $\Lambda_{\text{XRD}}$, using the equation\cite{yashar1999nanometer}
\begin{equation}
\sin(\theta_{\pm})=\sin(\theta_B)\pm\frac{m\lambda}{2\Lambda_{\text{XRD}}},
\label{Eq: diffraction angle}
\end{equation}
where $m$ and $\lambda$ are the order of the reflection and the wavelength of radiation (Cu-K$_{\alpha}$), respectively.
The $\Lambda_{\text{XRD}}$ values correlate well with $\Lambda_{\text{nom}}$ obtained from the calibrated deposition rates as well as $\Lambda_{\text{calc}}$ calculated from the number of interfaces and the total film thickness, $d_{\text{film}}$ (Tab.~\ref{Tab: architecture}). 
The main experimental peak positions, (111), (200), and (220), of the SL coating with $\Lambda=1.5\;\text{nm}$ coincide well with those of the Mo$_{0.75}$N/TaN, Mo$_{0.5}$N/TaN, and MoN$_{0.5}$/TaN structures proposed by our DFT calculations (Fig.~\ref{FIG: XRD} b).
Also the corresponding in-plane lattice parameters 4.33\;\AA\;(Mo$_{0.75}$N/TaN), 4.27\;\AA\;(Mo$_{0.5}$N/TaN), and 4.32\;\AA\; (MoN$_{0.5}$/TaN) are in line with the experimental value 4.29\;\AA.

\begin{table}[h!]
\caption{Deposition times of the single TaN and MoN layers correlated with the architecture of MoN/TaN SL coatings.}
\centering
\begin{tabular}{cc|cccc}
\hline
\hline
\multicolumn{2}{c}{Deposition} &
\multicolumn{4}{c}{Architecture} \\
 $t_{\text{Mo}}$\;[s] & $t_{\text{Ta}}$\;[s] &  $d_{\text{film}}\;[\mu\text{m}]$  & $\Lambda_{\text{nom}}$\;[nm] & $\Lambda_{\text{calc}}$\;[nm] & $\Lambda_{\text{XRD}}$\;[nm] \tabularnewline 
\hline
4.9 & 6.6 & 2.0 & 1.5 & 1.50 & 1.40   \tabularnewline
9.7 & 13.2 & 1.8 & 3.0 & 2.67 & 2.65  \tabularnewline 		    		    
19.5 & 26.4 & 1.7 & 6.0 & 5.22 & 5.10 \tabularnewline 
\hline
\hline
\end{tabular}
\label{Tab: architecture}
\end{table}

\begin{figure}[h!t!]
	\centering
    \includegraphics[width=8.5cm]{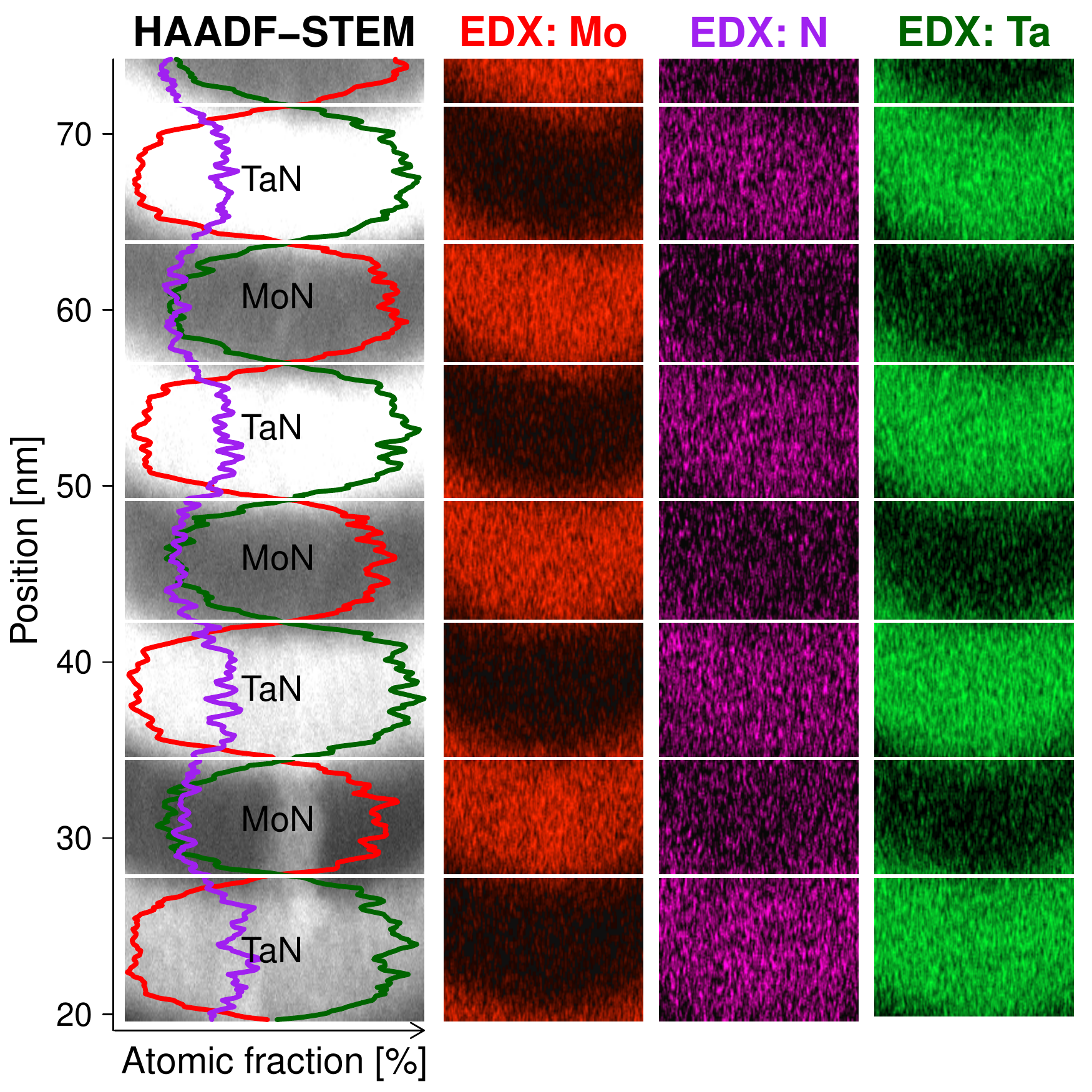}
    \caption{High-angle annular dark-field scanning transmission electron microscopy (HAADF-STEM) measurements for SL with $\Lambda=15\;\text{nm}$ together with Energy dispersive X-ray Spectroscopy (EDX) data for Mo, N, and Ta.}
\label{FIG: EDX}
\end{figure}

Despite the structural agreement with the {\it{ab initio}}-predicted candidates has been established, XRD results did not allow to estimate the actual coating composition.
To provide a clue on the vacancy type, we analysed chemistry of our superlattice coatings using EDX.
K$_{\alpha}$ and L$_{\alpha}$ peaks were selected for an analysis of N, Mo, and Ta content.
Fig.~\ref{FIG: EDX} depicts a representative semi-quantitative line-scan together with compositional maps.
Clearly, the N atomic fraction is significantly lower in the MoN layers in comparison with the TaN layers. 
This variance was not absolutely quantified due to a nearness of Ta (N$_{2,3}$) and Mo (M$_2$, M$_3$) peaks to the N-K$_{\alpha}$ peak in the energy spectrum. 
On the other hand, the intensities of Ta (N$_{2,3}$) and Mo (M$_2$, M$_3$) peaks are significantly lower in comparison to the N-K$_{\alpha}$ peak. 
Therefore, a systematic reduction of N-K$_{\alpha}$ signal is visible and allows to display this variation semi-quantitatively, which points towards N vacancies in MoN layers of the SL.
Besides that, we cannot exclude a certain concentration of Mo and/or Ta vacancies.

As a next step, we go back to the DFT results (Fig.~\ref{FIG: Ef}) and identify the structural candidates that may actually exist under our specific deposition conditions.
We thus limit ourselves to the relevant subspace of the theoretical ($\mu_{\text{N}}$,$\mu_{\text{Mo}}$,$\mu_{\text{Ta}}$)-dependent phase diagram, which is the closest approximate to the N$_2$ partial pressure 0.12\;Pa, temperature 663\;K, and Mo-to-Ta sputter yield 1.41.
The temperature and pressure dependence of $\mu_{\text{N}}$ is introduced following Ref.\cite{reuter2001composition} and using the reference values at 700\;K tabulated in Ref.\cite{Stull1971-tu}.
The experimental Mo-to-Ta sputter yield expresses the availability of the respective metal species in the deposition process and hence, can be related to the ratio of their chemical potentials, $\mu_{\text{Ta}}$/$\mu_{\text{Mo}}$.
The reason for taking $\mu_{\text{Ta}}$/$\mu_{\text{Mo}}$ instead of $\mu_{\text{Mo}}$/$\mu_{\text{Ta}}$ is that 0 refers to Mo(Ta)-rich conditions and the lower the $\mu_{\text{Mo}}$ ($\mu_{\text{Ta}}$) values are, the more we approach the Mo(Ta)-poor state.

\begin{figure*}[t]
	\centering
    \includegraphics[width=13cm]{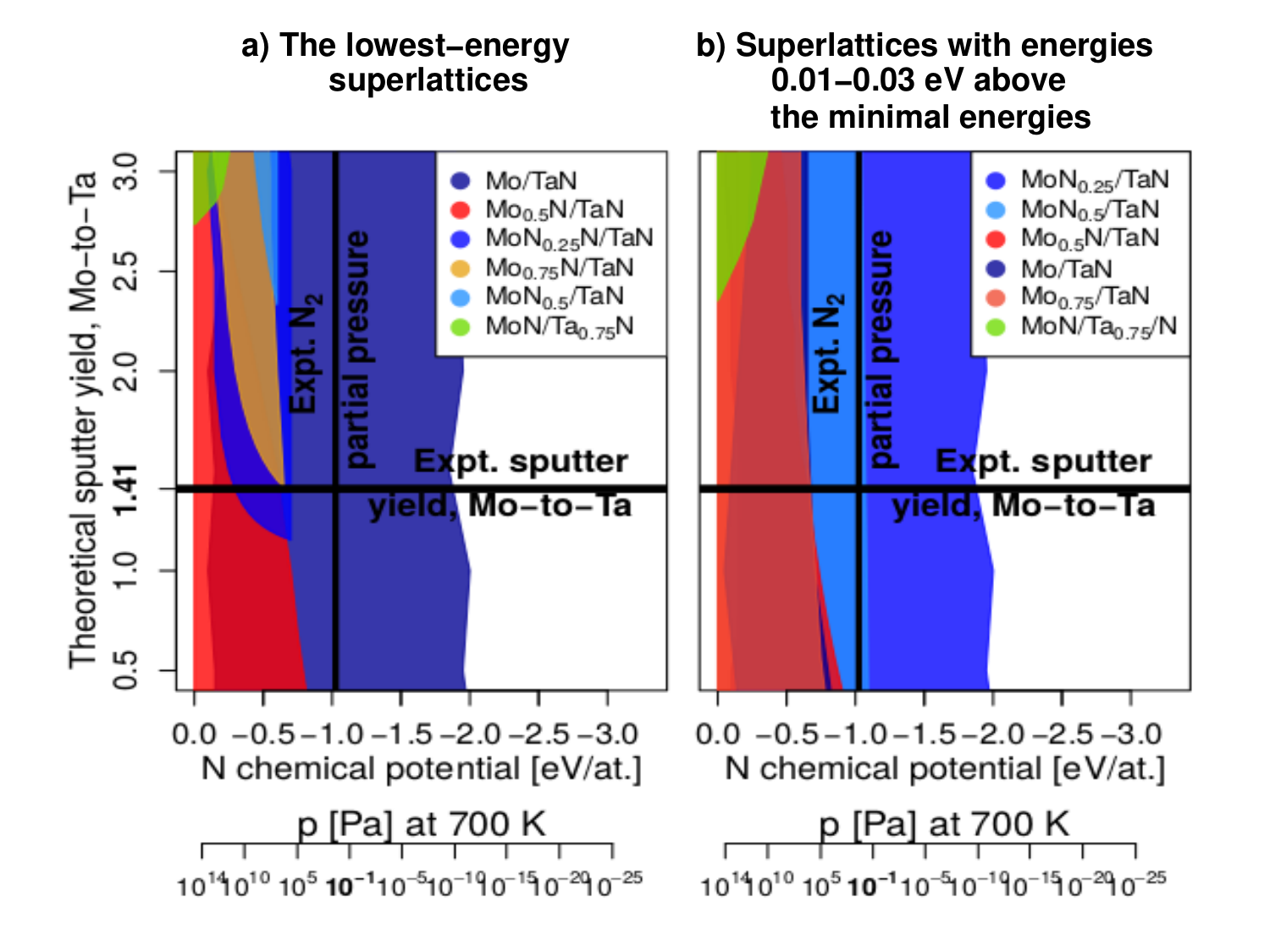}
	\caption{The most stable (a) and the low-energy metastable (b) MoN/TaN SLs as a function of N chemical potential and theoretical Mo-to-Ta sputter yield. The cross section of the two black solid lines---denoting the experimental N$_2$ partial pressure and sputter yield---suggests, which of the {\it{ab initio}} predicted candidates is the one produced during the deposition.}
\label{FIG: N2partialPressure}
\end{figure*}

Fig.~\ref{FIG: N2partialPressure} reveals that the intersection of the experimental N$_{2}$ partial pressure and Mo-to-Ta sputter yield falls within the stability range of Mo/TaN SL.
Despite supporting our hypothesis on N vacancies in MoN layers, the simulated XRD of Mo/TaN rather deviates from the experimental record.
Plotting stability ranges of the metastable states (with $E_f$ about 0.03\;eV/at. above the minimum energy), MoN$_{0.5}$/TaN SL is obtained as another structural candidate under our deposition conditions.
The good agreement between the experimental XRD, the in-plane lattice parameter and the corresponding DFT data for MoN$_{0.5}$/TaN indicate that this SL is probably close to the synthesised one.

\subsection{Elastic properties}\label{B:Elasticity}
As a next step, we study the impact of vacancies on the elasticity of MoN/TaN SLs and contrast the trends with these for the monolithic $\zeta\text{-Me}_x\text{N}_y$, Me$=$Mo, Ta.
Applying the {\it{ab initio}} stress-strain method, we obtained full elastic matrices $C_{ij}$ and estimated the mechanical stability of the corresponding system by calculating the minimal eigenvalue $\lambda_{\min}$ (Tab.~\ref{Tab: Cij}). 

Our results clearly underpin the important stabilisation role of vacancies in $\zeta\text{-MoN}$. 
While the defect-free $\zeta\text{-MoN}$ is mechanically unstable (especially due to the low $C_{44}$ leading to a negative $\lambda_{\text{min}}$), it becomes stable if either $12.5\text{--}25\%$ of Mo or $12.5\text{--}50\%$ of N vacancies is present.
Regarding $\zeta\text{-TaN}$, both Ta and N vacancies preserve mechanical stability as long as their concentration is below 25\;at.\%.
These findings suggest that not only the most stable Mo$_{0.88}$N and MoN$_{0.5}$ (as suggested by Fig.~\ref{Eq: Ef}), but also some metastable states (e.g., MoN$_{0.75}$) might be experimentally accessible.
On the other hand, the TaN$_{0.5}$ predicted as the most stable polymorph under the low N$_{2}$ partial pressure conditions might not be synthesisable as a consequence of its mechanical instability.

As reported earlier\cite{koutna2018stability}, the defect-free MoN/TaN SL with $\Lambda\approx 1.91\;\text{nm}$ (composed of the $\zeta$-phases), is mechanically stable.
Considering the instability of the $\zeta\text{-MoN}$ phase itself, this is a highly interesting result.
Based on Sec.~\ref{Sec1,ab initio}, we propose that stabilisation of MoN layers is obtained through the interface with the strong $\zeta\text{-TaN}$ and the imperfect structure of the $\zeta\text{-MoN}$ itself, which partially relaxes towards the mechanically stable $\omega\text{-MoN}$ phase (see Ref.~\cite{koutna2018stability}). 
Vacancies in MoN layers together with perfect TaN layers yield positive definite elastic matrices, i.e., Mo$_{x}$N$_y$/TaN SLs are mechanically stable.
This phenomenon relates to the increasing $C_{44}$, and hence, increasing resistance against shearing.

\begin{table*}[ht]
   \caption{Structural and elastic data for monolithic $\zeta\text{-phases}$ and SLs. Vacancy type and content, V$_{\text{Mo}}$, V$_{\text{Ta}}$, and V$_{\text{N}}$ (in \%), is correlated with structural parameters, $a$, $c$, and $\Lambda$ (in \AA), minimal eigenvalue of the elastic matrix, $\lambda_{\text{min}}$ (in GPa), ratio of the projections of elastic matrix onto tetragonal and cubic symmetries, $d_F^{\text{tetr/cub}}$, elastic constants, $C_{ij}$ (in GPa), elastic moduli, $B$, $G$, $E$ (in GPa), Pugh's and Poisson's ratios, $B/G$ and $\nu$.}
\label{Tab: Cij}		
    \vspace{0.5cm}
    \centering\small\setlength\tabcolsep{2pt}
        \hspace*{-2cm} 
\begin{tabular}{ccccccccccccccccccccc}
		\hline \hline
		$\zeta\text{-MoN}$ \\
		V$_{\text{Mo}}$	 & V$_{\text{N}}$ & $a$ & $c$ & $c/a$ & $\lambda_{\text{min}}$ & $d_F^{\text{tetr/cub}}$ & $C_{11}$ & $C_{12}$& $C_{13}$ & $C_{33}$ & $C_{44}$ & $C_{66}$ &  $B$ & $G$ & $E$ & $B/G$ & $\nu$ \\
		\hline
	0 & 0&  4.25  & 4.54 & 1.068  & $-99$ & 0 & 465  & 144 & 250 & 454 & $-99$ & 105 &   \\
    12.5 & 0 & 4.31 & 4.33 & 1.005 & 77 & 0.90 & 487  & 147 & 161  & 460 & 95& 83 & 264 & 114  &300  & 2.31  & 0.31 \\
	25 & 0 &  4.25  & 4.30 & 1.012 & 75 & 0.91 & 328  &157  & 138 & 346 & 79& 86  & 207 & 86 & 228 & 2.40 & 0.32\\
	50 & 0 & 4.46  & 4.42 & 0.992 & $-11$ & 0.93 & 60  & 20  & 45 & 65 & 40 & 41  \\
	0 &	12.5 & 4.23  & 4.59 & 1.085 & 59 & 0.24 & 495  & 217  & 199 & 189 & 87 & 115 & 228 & 84 &225  &2.70  &0.34 \\
    0 & 25 &  4.27  & 4.30 & 1.007 & 73 & 0.95 & 489  & 215  & 221 & 466 & 90 & 92 & 306 & 105 & 282  & 2.91  & 0.35 \\
	0 & 50 &  4.19  & 4.22 & 1.007 & 105 & 0.95 & 470 & 215  &201  & 497 & 108 & 117 & 306  &121  & 319 & 2.46 & 0.32 \\
		\hline
		$\zeta\text{-TaN}$ \\
		V$_{\text{Ta}}$	 & V$_{\text{N}}$ & $a$ & $c$ & $c/a$ &$\lambda_{\text{min}}$ & $d_F^{\text{tetr/cub}}$ & $C_{11}$ & $C_{12}$& $C_{13}$ & $C_{33}$ & $C_{44}$ & $C_{66}$ &  $B$ & $G$ & $E$ & $B/G$ & $\nu$ \\
		\hline
	0 & 0  & 4.20 & 5.12 & 1.219 & 106 & 0.03 &727  &  160 & 154 & 343 & 108 & 203 & 197 & 159 & 403 & 1.79 & 0.26 \\
	12.5 & 0  & 4.35  & 4.45 & 1.023 & 84 & 0.38 & 533  &145 & 234  & 309 & 106& 108 &  283 & 107  & 285  & 2.65  & 0.33 \\
	25 & 0 & 4.34  & 4.37 & 1.007 & 104 & 0.99 & 425 &136  & 134& 421 & 113& 107 & 161 & 57 & 153 & 2.84 & 0.34 \\
	0 & 12.5  & 4.20   & 5.07 & 1.207 & 99 & 0.11 & 592  & 211  & 161 & 290 & 100 & 177 & 264 & 131 &338  &2.00  &0.29 \\
	0 & 25 & 4.21  & 4.93 & 1.171 & 1 & 0.13 & 459 & 259  & 203 & 128 & 89 & 140 &  161 & 57 & 153 & 2.84 & 0.34 \\
	\hline
	MoN/TaN \\
		V$_{\text{Mo}}$	 & V$_{\text{N (MoN)}}$ & $a$ (in-plane) & $\Lambda$ & & $\lambda_{\text{min}}$ & $d_F^{\text{tetr/cub}}$ & $C_{11}$ & $C_{12}$& $C_{13}$ & $C_{33}$ & $C_{44}$ & $C_{66}$ & $B$ & $G$ & $E$ & $B/G$ & $\nu$ \\
		\hline	
		0 & 0 & 4.24 & 19.1 &  & 25 & 0.23 &  696  &  170 & 173 & 407 & 27 & 148 & 305 & 91 & 248 & 3.35 & 0.36 \\
	25 & 0  & 4.34 & 17.8 & & 133 & 0.30 &  596  & 131 & 153 & 486 & 91 & 115 &  283 & 133 &345 & 2.13 & 0.30 \\
		50 & 0  & 4.24 & 19.1 & & 122 & 0.11&  608  & 171 & 130 & 431 & 72 & 127 & 274 & 122 & 318 & 2.25 & 0.31 \\	
		0 & 12.5  & 4.23 & 18.9 & & 37 & 0.10 & 666 & 207 & 174 & 346 & 14 & 144 & 290 & 78 & 214  &  3.73 & 0.38\\
	0 & 25.0  & 4.23 & 18.7 & & 104 & 0.12 & 616 & 201 & 189 & 331  & 58 & 141 & 289 & 104 & 277 & 2.79 & 0.34 \\
		0 & 50.0  & 4.32 & 17.2 & & 160 & 0.04 & 668 & 152 & 169 & 514 & 123 & 129 & 317 & 160 & 411 & 1.98 &0.28  \\ 
	0 & 75.0  & 4.32 & 16.9 & & 150 & 0.04 & 641 & 136 & 179 & 463 & 108 & 153 & 301 & 150 & 386 & 2.01 & 0.29 \\
    0 & 100  & 4.39 & 15.6 & & 145 & 0.11 & 598 & 172 & 171 & 521 & 104 & 150 & 302 & 145 & 375 & 2.09 & 0.29 \\
	\hline
	\hline	
		\end{tabular}
        \hspace*{-2cm}
\end{table*}

While our pre-study\cite{koutna2018stability} showed that the cubic MoN and TaN transform towards the tetragonal $\zeta$ phases when no vacancies are present in the SL, Sec.~\ref{sectionA} suggested that $\zeta$-phases actually transform back to the cubic structure when vacancies are introduced in the simulation cell.
To support this hypothesis, we perform a simple analysis of elastic symmetry.
We note that elastic matrices, $\mathbb{C}$, calculated for defected systems generally constitute 21 independent elastic constants (due to the chemical disorder of vacancies and their high content).
The number of independent elastic constants decreases by projecting $\mathbb{C}$ onto a higher symmetry, given by the symmetry of a simulation cell before the relaxation (which corresponds to the macroscopic---experimental---symmetry).
Following \citet{moakher2006closest}, we can search for the best fitting projection, $\mathbb{C}_{\text{sym}}$, for our elastic tensor without any symmetry assumptions. 
Such approach consists in minimising the distance $d_F(\mathbb{C},\mathbb{C}_{\text{sym}})$ in the Frobenius (Euclidean) norm\cite{koutna2018stability}.
Here, the lowest symmetries, triclinic and monoclinic, were excluded from the analysis.
 
Evaluating $d_F(\mathbb{C}, \mathbb{C}_{\text{sym}})$ for cubic, tetragonal, hexagonal, and ortorhombic symmetry classes shows that the expected $\mathbb{C}_{\text{tetr}}$ indeed represents the closest projection for all systems. 
Nevertheless, examination of the ratio
\begin{equation}
d_F^{\text{tetr/cub}}:=d_F(\mathbb{C},\mathbb{C}_{\text{tetr}}) / d_F(\mathbb{C}, \mathbb{C}_{\text{cub}})
\end{equation}
reveals interesting facts in the case of defected $\zeta$-phases (Tab.~\ref{Tab: Cij}). 
As $d_F^{\text{tetr/cub}}$ approaches 1 for $\zeta\text{-Ta}_{0.75}\text{N}$, $\zeta\text{-Mo}_{0.87}\text{N}$, $\zeta\text{-Mo}_{0.75}\text{N}$, and $\zeta\text{-MoN}_{0.5}$, the $\mathbb{C}_{\text{cub}}$ and $\mathbb{C}_{\text{tetr}}$ are equally suitable for describing elasticity of these systems. 
On the other hand, the low $d_F^{\text{tetr/cub}}$ obtained for $\zeta\text{-TaN}_{0.87}$ ($\sim 0.15$) shows that the tetragonal symmetry of the elastic tensor is well preserved.
Such conclusions are in a perfect agreement with the discussion in Section~\ref{sectionA}.
Defected MoN/TaN SLs yield rather low $d_F(\mathbb{C},\mathbb{C}_{\text{tetr}}) / d_F(\mathbb{C}, \mathbb{C}_{\text{cub}})$ values, which justifies their overall tetragonal symmetry. 
In the case of the defect-free MoN/TaN, the more noticeable deviations from tetragonality can be ascribed to the relaxation from $\zeta\text{-MoN}$ towards the lower-symmetry $\omega\text{-MoN}$. 

Since the tetragonal projection was shown to be the best approximant of all our elastic tensors, we used it to evaluate the 6 independent elastic constants (Tab.~\ref{Tab: Cij}) as well as the corresponding polycrystalline bulk, $B$, shear, $G$, and Young's, $E$, moduli (Fig.~\ref{FIG: B,G,E}).
The overall highest $B$, $G$, and $E$ values (306, 121, and 319\;GPa, respectively) in the Mo--N system are obtained for $\zeta\text{-MoN}_{0.5}$.  
Besides, N-deficient $\zeta\text{-MoN}_{0.89}$, $\zeta\text{-MoN}_{0.75}$, and $\zeta\text{-MoN}_{0.5}$ show more ductile character (based on the Pugh's\cite{pugh1954xcii} and Frantsevich's criteria\cite{frantsevich1983elastic}) than their Mo-deficient counterparts, $\zeta\text{-Mo}_{0.89}\text{N}$ and $\zeta\text{-Mo}_{0.75}\text{N}$.
Defects in $\zeta\text{-TaN}$ cause a significant drop in $B$, $G$, and $E$ moduli: from 286, 159, and 403\;GPa (perfect $\zeta\text{-TaN}$) down to 161, 57, and 153\;GPa ($\zeta\text{-TaN}_{0.75}$), respectively.
Bulk moduli of the N-deficient SLs vary between 289--317\;GPa, while they fall down from 305\;GPa (defect-free system) to 274\;GPa when 50\% of Mo sites is unoccupied.
Importantly, some of the $B$, $G$, and $E$ values---e.g., 317, 160, and 411\;GPa of the MoN$_{0.5}$/TaN---exceed those of the corresponding binaries, $\zeta\text{-Me}_{x}\text{N}_y$, Me$=$Mo, Ta. 
Moreover, Mo$_x$N$_y$/TaN SLs show promising ductile character, based on their relatively high $B/G$ and $\nu$ values (Tab.~\ref{Tab: Cij}).

\begin{figure}[h!]
	\centering
    \includegraphics[width=8.5cm]{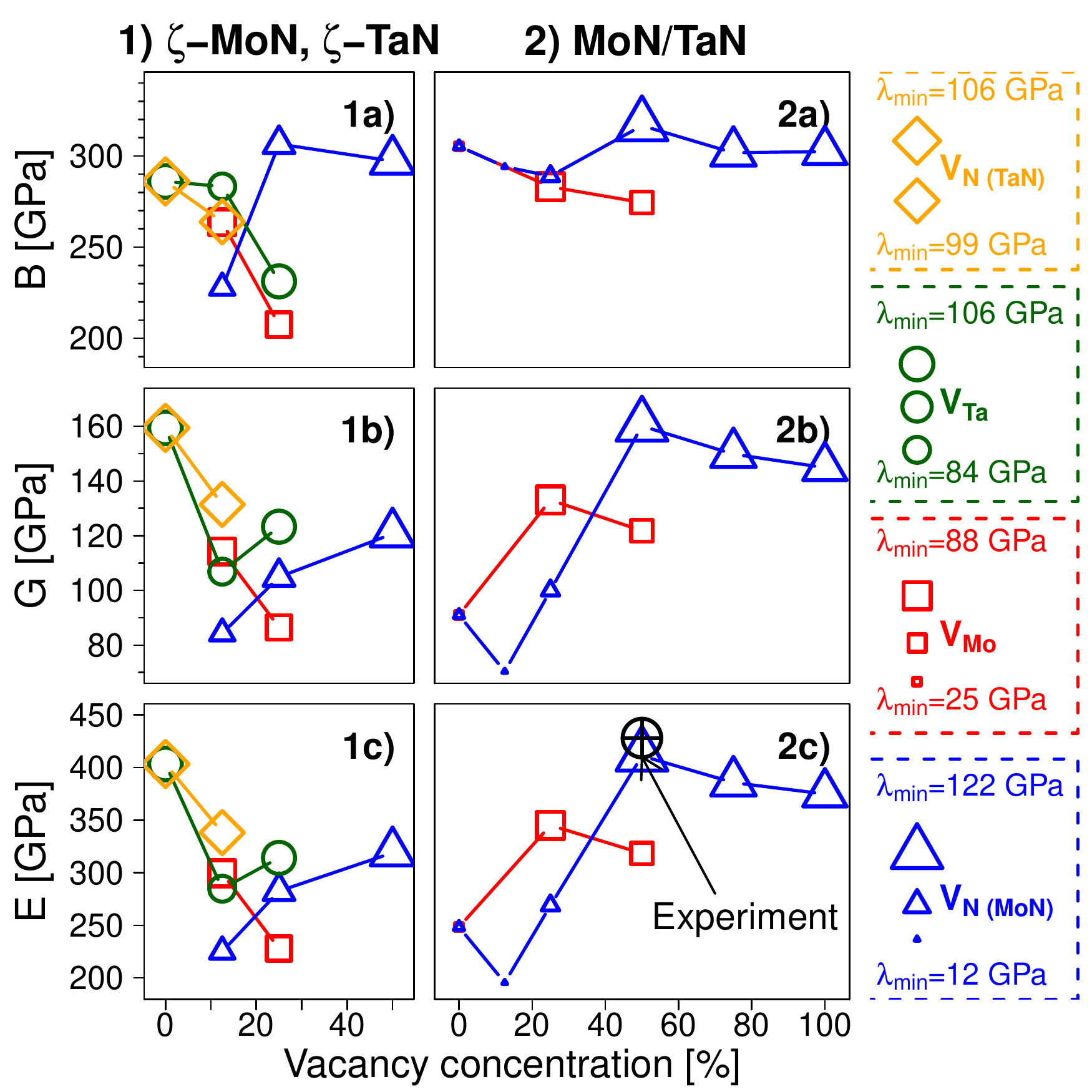}
	\caption{The polycrystalline bulk, $B$, shear, $G$, and Young's moduli, $E$, of the single-phase (1a--c) and the SL systems (2a--c) as functions of the vacancy content. The symbol size scales with the minimal eigenvalue, $\lambda_{\text{min}}$, hence, rating mechanical stability of the corresponding structures.}
\label{FIG: B,G,E}
\end{figure}

In line with the DFT predictions, also our experiments show that the MoN/TaN SL coatings have superior mechanical properties (Tab.~\ref{Tab: mechanical properties}).
The indentation hardness is found within the range of 31--34\;GPa for bilayer periods 1.4, 2.65, and 5.1\;nm.
We note that due to the preferred orientation, coatings on MgO are slightly harder.
The indentation modulus shows a slight dependence on $\Lambda$: $E$ decreases from the 428\;GPa ($\Lambda\sim 1.40\;$\;nm) down to 379\;GPa ($\Lambda\sim 5.10$\;nm). 
Tab.~\ref{Tab: mechanical properties} justifies that such evolution is not caused by residual stresses\cite{music2017correlative} that vary only slightly (between 1.7 and 2.1\;GPa).
The experimentally measured indentation value of $(428\pm23)\;\text{GPa}$ for the 1.4\;nm SL, perfectly agrees with the {\it{ab initio}} calculated 411\;GPa of MoN$_{0.5}$/TaN, thus supporting the hypothesis on the chemistry of our SL coatings.  

\begin{table}[h!t!]
\caption{Young's moduli, $E$, hardnesses, $H$, and residual stresses, $\sigma_{\text{r}}$, of our SL coatings (deposited on Si or MgO substrate) correlated with bilayer period, $\Lambda_{\text{XRD}}$.}
\centering
\begin{tabular}{c|cccc}
\hline
\hline
\multicolumn{1}{c}{Architecture} &
\multicolumn{4}{c}{Mechanical properties} \tabularnewline
 $\Lambda_{\text{XRD}}$\;[nm] & $E$\;[GPa] & $H_{\text{Si}}$\;[GPa] & $H_{\text{MgO}}$\;[GPa] & $\sigma_{\text{r, Si}}$\;[GPa]\tabularnewline 
\hline
 1.40   & $428\pm23$ & 31.0$\pm$1.6 & 32.3$\pm$1.3 & 1.8$\pm$0.1  \tabularnewline
 2.65   & 388$\pm$21 &  31.8$\pm$1.6 & - & 2.1$\pm$0.1 			 \tabularnewline 		    		    
 5.10   & 379$\pm$14 & 31.2$\pm$1.6 & 33.5$\pm$1.3 & 1.7$\pm$0.1  \tabularnewline 
\hline
\hline
\end{tabular}
\label{Tab: mechanical properties}
\end{table}

\subsection{Dynamical stability and electronic structure}\label{C:Phonons}
To complete the picture of MoN/TaN SLs as well as to underpin our hypothesis on the presence of N vacancies in MoN layers, we calculate vibrational and electronic properties.

\begin{figure}[h!t!]
	\centering
    \includegraphics[width=8cm]{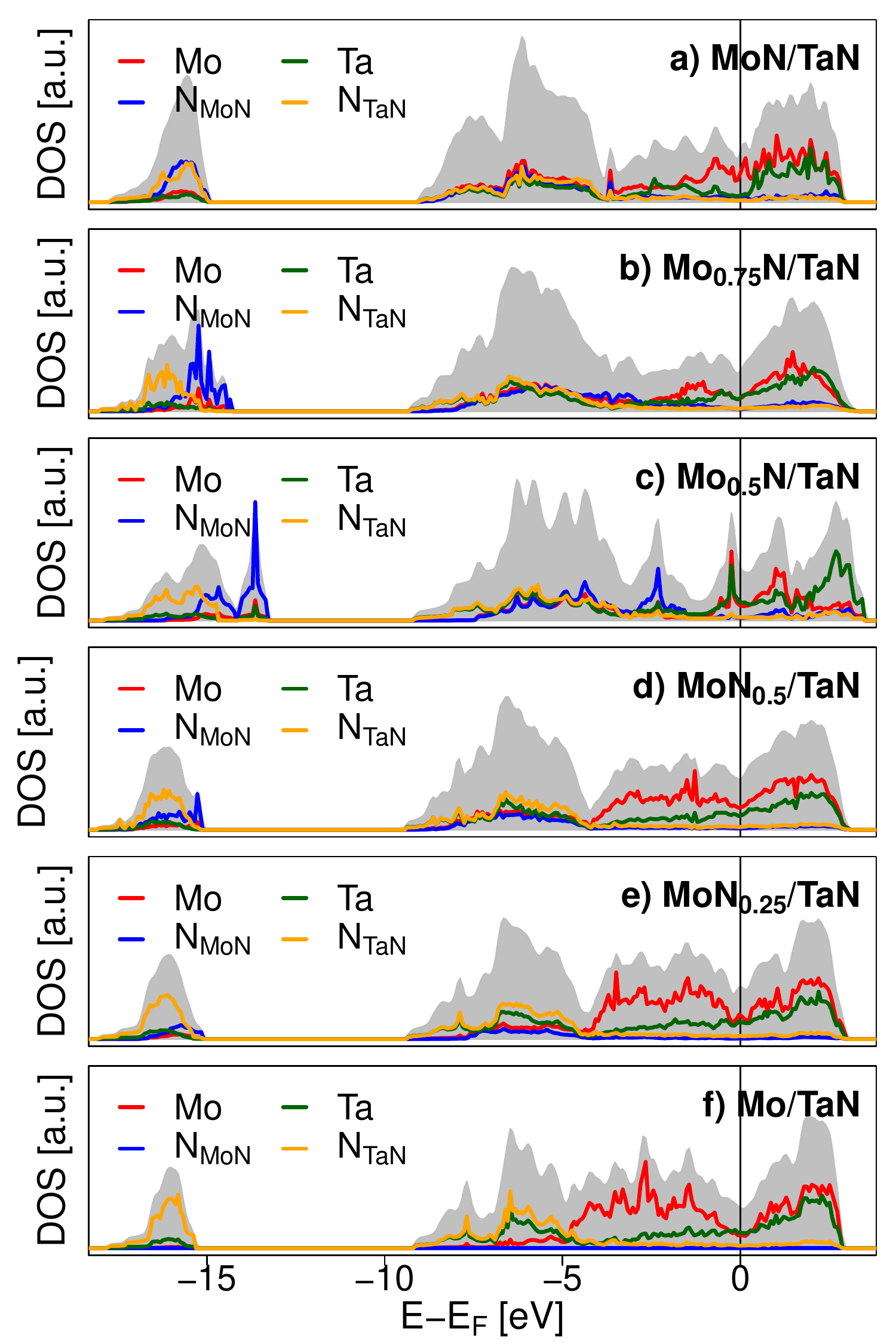}
    \caption{Electronic density of states (DOS) for low-energy SLs. The grey-shaded areas show the total DOS, while the red, green, blue, and yellow lines correspond to the partial contribution from Mo, Ta, and N orbitals in MoN and TaN, respectively.}
\label{FIG: DOS}
\end{figure}

Starting with the electronic structure, Fig.~\ref{FIG: DOS} indicates that the metallic character of the SL is preserved, regardless the vacancy content.
Local character of the total DOS close to the Fermi level, $E_F$, is in line with our previous findings on mechanical stability.
In the case of defect-free MoN/TaN, the vicinity of $E_F$ is dominated by $\text{Mo-d}$ states, while the contribution from $\text{Ta-d}$ orbitals is rather small.
The Fermi energy is also off the nearest minimum of the total DOS, which is in line with its mechanical instability.
A reduction of Mo-d contribution close to $E_F$ is accompanied by an increase of Ta-d states as well as the interstitial DOS.
The latter is a consequence of numerous unoccupied lattice sites and hence, broken bonds of the neighbouring atoms. 
For N-deficient SLs, Mo-d states dominate especially in the range of $-4\;\text{eV}$ up to $E_F$.  
Additionally, two sharp N-derived peaks develop for Mo$_{0.5}$N/TaN.
As intuitively expected, the DOS profile of the bcc-Mo containing SL, Mo/TaN, largely deviates from the defect-free case.
  
\begin{figure}[h!t!]
	\centering
    \includegraphics[width=7.5cm]{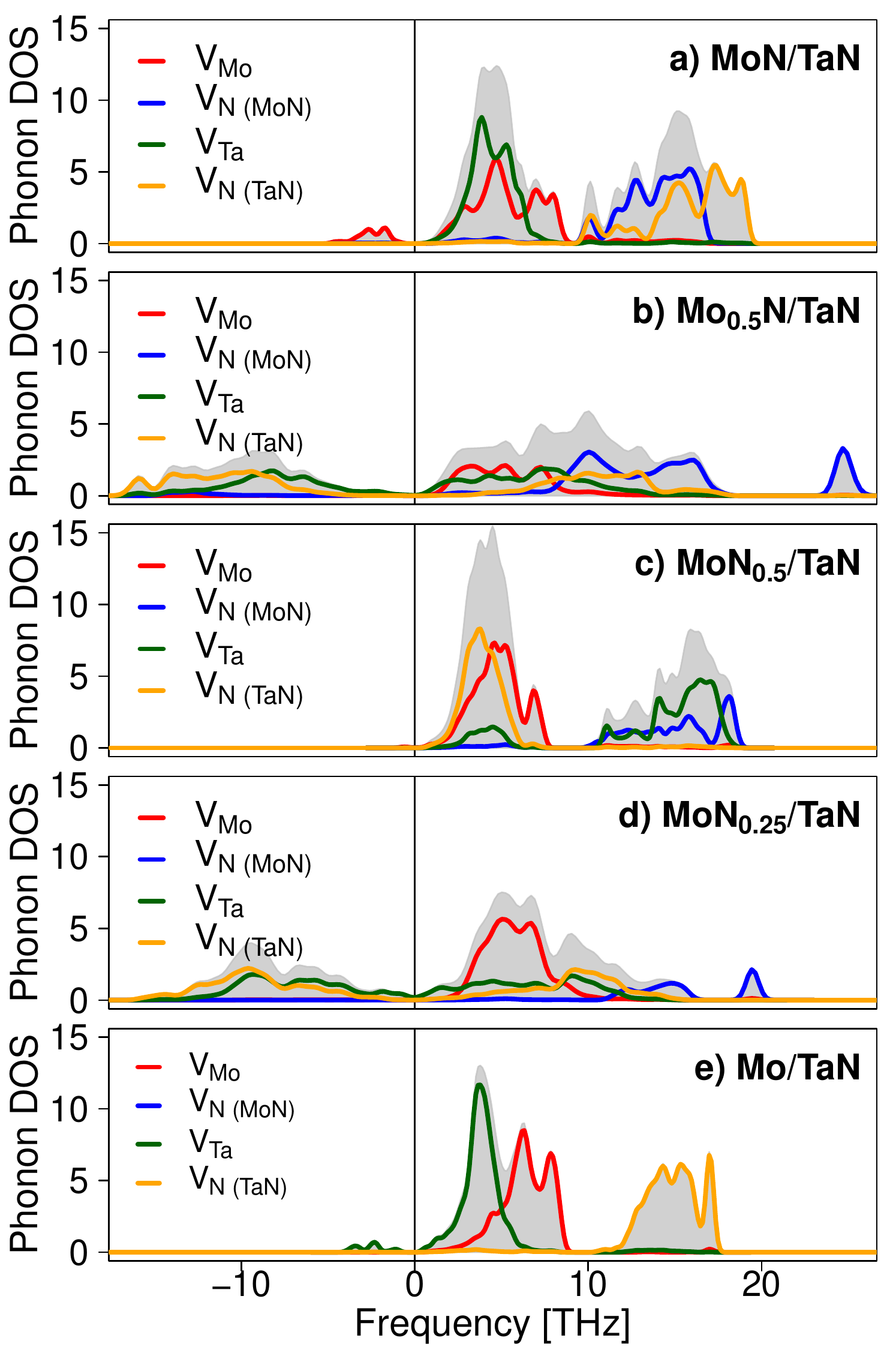}
	\caption{Phonon density of states (DOS), in states/THz/at., for low-energy SLs. The grey-shaded areas show the total DOS, while the red, green, blue, and yellow lines correspond to the partial contribution from Mo, Ta, and N atoms in MoN and TaN, respectively.}
\label{FIG: phonons}
\end{figure}

Furthermore, Fig. \ref{FIG: phonons} uncovers dynamical instability of the defect-free SL due to the presence of imaginary phonon frequencies (caused by vibrations of the Mo atoms).
The dynamical stability of the $\zeta\text{-TaN}$ thus does not compensate for the instabilities of the $\zeta\text{-MoN}$ (or $\zeta/\omega\text{-MoN}$).  
The Mo$_{0.5}$N/TaN and MoN$_{0.25}$/TaN SLs yield even higher phonon density of states in the imaginary frequency range, mainly originating from vibrations in TaN layers.
As already mentioned, the high vacancy content in MoN causes a compression of $\zeta\text{-TaN}$ layers along the $z$-axis (cf.~Fig.~\ref{FIG: lattice parameters}), which is accompanied by several soft modes in phonon spectra.   
For the same reason, vibrations of Ta atoms induce some minor instabilities of Mo/TaN structure, though no imaginary phonons frequencies are present in bcc-Mo.  
Importantly, all imaginary phonon frequencies are eliminated in MoN$_{0.5}$/TaN SL.
This vacancy content is thus sufficient to stabilise MoN layers, but not too high to destabilise TaN layers.   	

According to DFT calculations, the MoN$_{0.5}$/TaN SL---supposedly a close approximant to the experimental coatings---does not only exhibit a local DOS minimum at the Fermi level (which is a sign for stability), but actually is the only vibrationally stable configuration.
Nevertheless, we note that the here presented phonon calculations are far from being an accurate description of the phonon properties as a function of defect type, concentration, and configuration.
A careful supercell size optimisation and/or different defect distribution may lead to a dynamical stabilisation of other defected systems, not only MoN$_{0.5}$/TaN.

\section{Summary and conclusions}
Structure-stability-elasticity relations for cubic-based MoN/TaN superlattices were established by modelling and experimental techniques. 
Our material system presented a particular versatility in physical properties, as both MoN and TaN have a strong affinity for vacancies and can easily structurally transform due to the presence of interfaces (and the corresponding non-homogeneities of the electronic charge density).

Quantum-mechanical DFT calculations identified the most energetically favourable SL candidates, Mo$_{0.5\text{--}0.75}$N/TaN, Mo/TaN, MoN$_{0.25-0.5}$/TaN, and MoN/Ta$_{0.5\text{--0.75}}$N, depending on various deposition conditions, i.e., on the values of chemical potentials.
SLs with vacancies in MoN layers clearly dominated, suggesting that a high Mo or N deficiency is expectable for experimental coatings.
Indeed, the simulated XRD patterns of the Mo$_{0.5\text{--}0.75}$N/TaN and MoN$_{0.5}$N/TaN fitted perfectly to the measured XRDs for magnetron sputter-deposited MoN/TaN SLs with $\Lambda\approx1.5\text{--}6$\;nm.
Chemical investigations using EDX further suggested that our SL coatings contain N vacancies.
Additional DFT analysis pointed towards MoN$_{0.5}$/TaN as the most likely structural variant under our deposition conditions.

Calculations of elastic properties proved a stabilisation effect of vacancies in terms of mechanical stability and suggested an improved ductility/toughness of MoN/TaN SLs as compared to the monolithic phases as well as transition metal nitride systems in general.
The {\it{ab initio}} polycrystalline Young's modulus of MoN$_{0.5}$/TaN (411\;GPa) perfectly agreed with the experimental indentation modulus (428$\pm$23\;GPa), thus supporting our hypothesis on N vacancies in MoN layers of SL coatings.
The measured indentation hardness reached up to 31--34\;GPa.
Phonon calculations further revealed that the MoN$_{0.5}$/TaN structural candidate is the only one that is vibrationally stable.

Our complex analysis of vacancy-stabilised MoN/TaN SLs underlines the high predictive power of modelling as well as the necessary symbiosis between theory and experiment in order to design novel materials.

\section*{Acknowledgements}
NK acknowledges the DOC fellowship by the Austrian Academy of Sciences, \"{O}AW. 
DH and MB highly appreciate the support by the Austrian Science Fund, FWF, (P 30341-N36).
JZ acknowledges the CEITEC Nano Research Infrastructure (ID LM2015041), funded by the Ministry of Education, Youth and Sports of the Czech Republic, MEYS CR.
M\v{S} acknowledges the Project CEITEC 2020 (LQ1601).
MF acknowledges the research infrastructure IPMINFRA (LM2015069) and the Czech Academy of Sciences through the Fellowship of J. E. Purkyn{\v e}.
MF and M\v{S} acknowledge the Czech Science Foundation (GA 16-24711S) the Institutional Project (RVO:68081723).

Computational resources were provided by the Vienna Scientific Cluster (VSC), by MEYS CR under projects CESNET (LM2015042), CERIT-Scientific Cloud (LM2015085), and IT4Innovations National Supercomputer Center (LM2015070) within the program Projects of Large Research, Development and Innovations Infrastructures.

\newpage
\bibliography{biblio}

\end{document}